\documentclass[twocolumn,aps,pra,floatfix]{revtex4}
\usepackage{graphicx}
\usepackage[english]{babel}
\usepackage{amsmath}
\usepackage{amssymb}
\usepackage{ulem}
\usepackage[usenames,dvipsnames]{color}
\usepackage{color}
\definecolor{Blue}{rgb}{0.3,0.3,0.9}
\definecolor{orange}{rgb}{1,0.5,0}
\usepackage{subfigure}
\usepackage{bm}
\usepackage{titlesec}
\newcommand{\bbt}{\mathbb{T}}
\newcommand{\bbr}{\mathbb{R}}
\newcommand{\bbi}{\mathbb{I}}
\newcommand{\bbs}{\mathbb{S}}
\newcommand{\bbf}{\mathbb{F}}
\newcommand{\bbu}{\mathbb{U}}
\newcommand{\fract}[2]{\ensuremath{\tiny{\frac{#1}{#2}}}}
\newcommand{\bfk}{\mathbf k}

\begin{document}
\title{Implications of the  Babinet Principle for Casimir Interactions}
\author{Mohammad F. Maghrebi}
\affiliation{Center for Theoretical Physics, and Department of Physics, Massachusetts Institute of Technology, Cambridge, MA 02139, USA}
\author{Ronen Abravanel}
\affiliation{Department of Physics, Technion, Haifa 32000, Israel}
\author{Robert L. Jaffe}
\affiliation{Center for Theoretical Physics, and Department of Physics, Massachusetts Institute of Technology, Cambridge, MA 02139, USA}

\begin{abstract}
We formulate the Babinet  Principle (BP) as a relation between the scattering amplitudes for electromagnetic waves, and combine it with  {multiple scattering techniques} to derive new properties of Casimir forces.  We show that the Casimir force exerted by a planar conductor or dielectric on a \emph{self-complementary} perforated  planar mirror is approximately half that on a uniform mirror independent of the distance between them.  The BP suggests that Casimir edge effects are anomalously small, supporting results obtained earlier in special cases.  Finally, we illustrate how the BP can be used to estimate Casimir forces between perforated planar mirrors.

\end{abstract}
\maketitle
In 1948, Casimir predicted an attractive force between metal plates arising from {vacuum fluctuations of the electromagnetic field} \cite{Casimir48-2}.
A systematic {understanding} of the Casimir effect is important for a vast range of physical problems from high energy physics to condensed matter systems \cite{Milton01}.
The advent of precision experimental measurements of Casimir forces
\cite{Lamoreaux97, Mohideen98,Harris00,Bressi02} and the possibility of applications to micron- and nano-scale electromechanical devices has stimulated interest in developing   efficient ways to compute these forces both analytically \cite{Milton01, Kenneth06, Emig07, Bordag09} and numerically \cite{Gies06, Reid09, Pasquali09}.
In particular, a multipole scattering method has been developed and used to compute Casimir forces among multiple objects of {various} shapes and electromagnetic properties \cite{Emig07,Rahi09}.
The essential ingredient in this formalism is the amplitude, expressed in a convenient basis, for electromagnetic waves to scatter from the individual objects.   While the conceptual foundations of the method harken back to earlier formalisms \cite{Balian77}, the successful implementation is quite recent.

The classical Babinet Principle relates the diffraction patterns of waves scattering from two \emph{complementary screens}, taken to be flat surfaces, the holes in one being filled in the other and {\it vice versa\/}.  The screens are assumed to have negligible thickness and to enforce boundary conditions on the scattering field, either Dirichlet (D), or Neumann (N) {for scalar fields}, or perfectly conducting (EM)  {for electromagnetic waves}.  The conflicting assumptions of perfect {conductivity} and negligible thickness place restrictions on the conditions where the BP can be applied.  At the end we estimate these conditions for a good conductor like gold.

We first state the Babinet Principle in a form suited to our purposes, as a relation between scattering amplitudes.  Then we show how the BP can be combined with the scattering theory approach to Casimir forces to make predictions for physically interesting configurations.   We start with self-complementary geometries in which the screen $\Sigma$ and its complement, $\widetilde\Sigma$, are identical up to a translation or rotation.  In this case we use the BP to show that the electromagnetic Casimir force between such a screen and a perfect mirror is approximately half the force between two perfect mirrors.  Subject to a proviso about the locality of Casimir forces, we suggest that Casimir edge effects are anomalously small in electromagnetism.  Finally we show how the BP can be used to compute the Casimir force between perforated screens.

\begin{figure}[h]
 \includegraphics[scale=1]{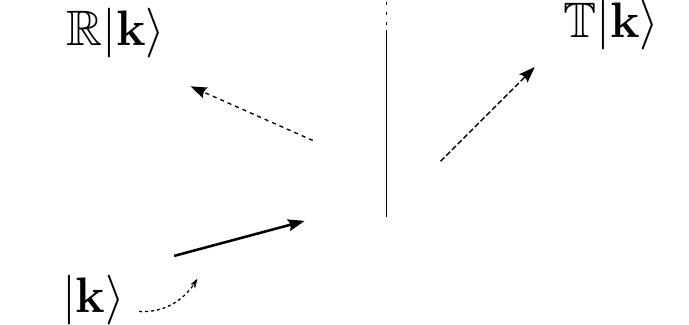}
  \caption{The two different channels of the scattering matrix. $\bbr$ characterizes the scattering back to the same side as the incident field while $\bbt$ gives the forward scattering to the opposite side of the screen. } \label{fig1}
\end{figure}

The scattering of waves from a screen is described in terms of a scattering amplitude, $\bbf$, related to the unitary $S$-matrix by $\bbf=\fract{1}{2}\left(\bbs-\bbi\right)$.    In the absence of the screen $\bbf$ vanishes.  It is convenient to distinguish the contribution to $\bbf$ corresponding to transmission of waves  across a screen, denoted by $\bbt$, from the contribution corresponding to waves that are reflected, denoted by $\bbr$.  Thus, a  wave $|\phi_{\rm inc}\rangle$ incident from the left on a screen $\Sigma$ gives rise to a reflected wave $\bbr|\phi_{\rm inc}\rangle$ and a transmitted wave, $\bbt|\phi_{\rm inc}\rangle$, a situation displayed in Fig.~\ref{fig1}.

 For the sake of simplicity and clarity, we discuss the case of a scalar field and then quote the generalization to electromagnetism. Without loss of generality, we assume a planar incident wave $|\mathbf k\rangle$  impinging on the screen from the left side, $k_{z}>0$.   The scattering off the screen $\Sigma$ is defined by the ansatz
\begin{equation}
  \left|\phi\right\rangle= \begin{cases}
    \left|\mathbf k\right\rangle + \sum_{\mathbf k'}\bbr_{ \, \mathbf k',\mathbf k}\left|\mathbf k'\right\rangle & \hskip .1 in \mbox{on the left}\,,\\
          \left|\mathbf k\right\rangle + \sum_{\mathbf k'}\bbt_{\,\mathbf k',\mathbf k}\left|\mathbf k'\right\rangle & \hskip .1 in \mbox{on the right}\,.\\
  \end{cases}
\end{equation}
Note that the trivial forward scattering is separated out from the diffraction described by $\bbr$ and $\bbt$. If the same incident wave, $\left|\mathbf k\right\rangle$, shines on the complementary screen, $\widetilde \Sigma$, the scattering can be described by a similar ansatz with {the} corresponding scattering amplitudes denoted by $\widetilde\bbr$ and $\widetilde\bbt$. The scattering is determined by the boundary conditions that the scalar field obeys on the screen, which we take to be either Dirichlet or Neumann.  Applied to scalar fields,  the BP relates the diffraction of a field that obeys a Dirichlet boundary condition on $\Sigma$, to the diffraction experienced by a field that obeys a Neumann boundary condition on $\widetilde\Sigma$, and {\it vice versa\/}.

The proof of the Babinet Principle is discussed in many textbooks where it is usually assumed that one screen has compact support, while its complement extends to infinity.  We have extended the BP for both scalar and electromagnetic fields to the case where neither screen has compact support, {\it e.g.\/} each is a half-plate.  The proof is inspired by an argument in Ref.~\cite{Lifshitz-ECM}, and is based on the fact that a general solution to the wave equation can be cast {as a sum of terms even or odd in $z$.  In the Dirichlet (Neumann) case the odd (even) solution is trivial.  It is not hard to show that an odd solution in the Neumann case for screen $\Sigma$ can be constructed from the even solution for $\widetilde\Sigma$ in the Dirichlet case and {\it vice versa\/}. Uniqueness of solutions to the Helmholtz equation fixes the correspondence of solutions to be one-to-one.} {A linear combination of even and odd terms sets up the scattering ansatz with incoming wave only on the left side of the screen, and is the unique solution of the scattering problem.} The result is a relation between transmission and reflection matrices for the screen and its complement.  For the scalar case,
\begin{equation}\label{Eq. T-LL Scalar}
  \bbr_{ \,  \mathbf k',\mathbf k}^{D/N }-  \widetilde \bbr^{N/D}_{ \,  \mathbf k',\mathbf k} =\mp \mathbb I_{\mathbf k',\mathbf k}\,.
\end{equation}
where $\bbi_{\mathbf k' ,\mathbf k }\equiv (2\pi)^{2}\delta(\mathbf k_{\|}'-\mathbf k_{\|})$  with $\mathbf k_\|$ being the component of the planar wave parallel to the screen.  As a check, if the screen $\Sigma$ is the entire plane so the screen $\widetilde\Sigma$ is the null set, then $\bbr^{D/N}_{\bfk',\bfk}=\mp\bbi_{\bfk',\bfk}$ as expected.
For the other channel, the BP dictates
\begin{equation}
\label{Eq. T-RL Scalar}
  \bbt_{\,\mathbf k',\mathbf k}^{D/N }+  \widetilde \bbt^{N/D}_{ \,  \mathbf k',\mathbf k} = -\mathbb I_{\mathbf k',\mathbf k}\,.
\end{equation}
Once again this can be checked in the limiting case where $\Sigma$ is either a full screen or the null set.  For example in the former case $\bbt^{D/N}_{\bfk',\bfk}=-\bbi_{\bfk',\bfk}$, which cancels the incident wave on the left side of the  screen.

Similar equations hold for electromagnetism. However, a slight complication arises since the scattering from a screen generally mixes electric (E) and magnetic (M) polarizations. The Babinet Principle must then be cast in a matrix form,
\begin{align}
  \begin{pmatrix}
    \bbr^{ M M } -\widetilde \bbr^{ E E }&\hskip 0.1in \bbr^{ M E }+\widetilde \bbr^{  E M } \\
      \\ \bbr^{ E M }  +\widetilde \bbr^{ M E }& \hskip 0.1in\bbr^{ E E }-\widetilde \bbr^{ MM }
  \end{pmatrix}
    & = \mathbb I \large\begin{pmatrix}
   -1 &   0\\ 0 &  1
  \end{pmatrix},  \label{Eq. T-LL EM}\\
  \begin{pmatrix}
    \bbt^{ M M } + \widetilde \bbt^{ E E }& \hskip 0.1in \bbt^{ M E }-\widetilde \bbt^{  E M } \\
      \\ \bbt^{ E M }-\widetilde  \bbt^{ M E } &\hskip 0.1in \bbt^{ E E }+\widetilde \bbt^{ MM }
  \end{pmatrix}
    & = \mathbb I \large\begin{pmatrix}
     -1  &   0 \\ 0 & -1
  \end{pmatrix}, \label{Eq. T-RL EM}
  \end{align}
where the labels $\{\bfk,\bfk'\}$ have been suppressed.

\begin{figure}[h]
 \includegraphics[scale=1.1]{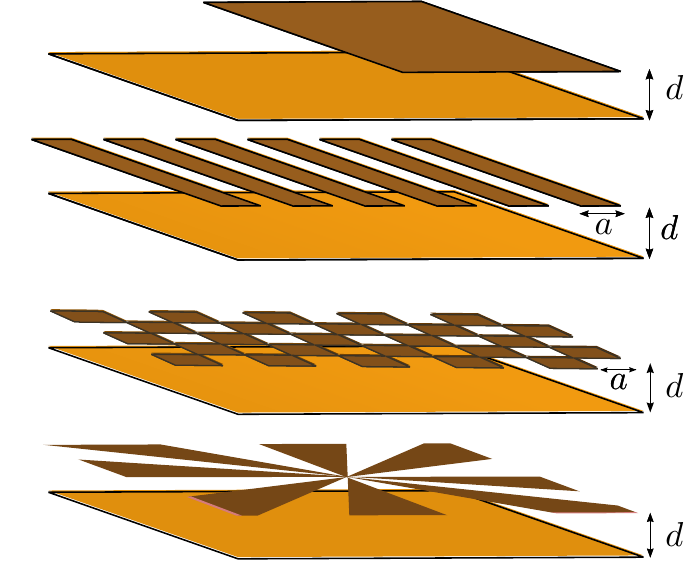}
 \caption{Self-complementary geometries, each facing an infinite plate. } \label{Fig-SelfDual}
\end{figure}

The Babinet Principle can be used to learn electromagnetic properties of new geometries based on their complementary partners. We can exploit this for geometries where we understand one side of {the} complementarity,  but it can be used even when the scattering properties of neither side are known, namely when the screens are \emph{self-complementary}, {\it i.e.} $\Sigma =\widetilde\Sigma$, up to a translation or rotation.   An example of such a geometry is a semi-infinite plate, whose complement is also a half-plate, infinitely extended in the opposite direction.  Other examples include a regular array of strips (with their size being the same as the gap between the strips), a checkerboard (an array of diagonally placed squares), and various angular subdivisions of the plane,  see Fig.~\ref{Fig-SelfDual}. Since these geometries are self-complementary, their scattering matrices are equal to their complement{'}s.

To incorporate the BP into the Casimir interaction, we take advantage of an expansion of the energy in multiple scatterings between objects \footnote{{The BP was used to relate the Casimir energy of an isolated screen to its complement in N. Graham and K. D. Olum, Phys. Rev. D {\bf 72}, 025013
(2005).}}. The Casimir interaction energy of two objects is given by $\mathcal E =\frac{\hbar c}{2\pi} \int_{0}^{\infty} d\kappa\, {\rm tr}\ln\left(\bbi- \bbf_1 \, \bbu_{12}\bbf_2\, \bbu_{21}\right)$ where the $\bbf$s are scattering amplitudes and the $\bbu$s are  translation matrices which capture the appropriate translations and rotations between the scattering bases for each object \cite{Kenneth06, Emig07}. When expanded in $ \bbf_1 \, \bbu_{12}\bbf_2\, \bbu_{21}$, this expression gives a series in multiple reflections which converges very rapidly in all cases that have been studied \cite{Scardicchio:2004fy, Maghrebi10, Maghrebi10-2}. In the case of parallel plates, for example, the first reflection captures about 93\% of the total energy, and in other cases like sphere-plate, it does even better.  We therefore focus on the first reflection term, $\mathcal E_{1}\equiv-\frac{\hbar c}{2\pi} \int_{0}^{\infty} d\kappa\, {\rm tr}\left(\bbf_1 \, \bbu_{12}\bbf_2\, \bbu_{21}\right)$, where we can directly apply the BP. Below we argue that corrections due to higher reflections are at most a few percent.

{Let us} consider an infinite plate, either dielectric or conductor,  opposite a conducting, self-complementary screen, some examples of which are given in Fig.~\ref{Fig-SelfDual}. Only the reflection amplitudes come into play in this configuration.  Reflection from an infinite plate, $\bbr_0^{P P'}$, is diagonal in polarization, hence $\bbr_0^{P' P}=\bbr_0^{P P} \delta_{P' P}$. So are the translation matrices $\bbu^{P' P}=\bbu^{PP}\delta_{P'P}$. The Casimir energy in the first reflection then becomes
\begin{equation}
\mathcal E_1 = -\frac{\hbar c}{2\pi} \sum_{P=E,M}\int_{0}^{\infty} d\kappa\, {\rm tr}\left(\bbr_0^{PP} \, \bbu^{PP}\bbr^{PP}\, {\bbu^{PP}}^\dagger\right)
\end{equation}
where $\bbr$ is the reflection amplitude for the self-complementary geometry.    Equation~(\ref{Eq. T-LL EM}) relates the reflection matrices of opposite polarizations, $\bbr^{MM} -\widetilde \bbr^{EE}=-\mathbb I$ and $\bbr^{EE} -\widetilde \bbr^{MM}=\mathbb I$.
Using this, one can show that the sum $\mathcal E_1+\tilde{\mathcal E}_1$ is equal to the energy of two infinite plates in the first reflection.
However, for a self-complementary geometry $\bbr=\widetilde\bbr$ up to a trivial translation or rotation matrix.  {${\mathcal E}_{1}$ and $\widetilde{\mathcal E}_{1}$ are therefore equal and both} equal to half the interaction of two infinite plates {in the first reflection approximation.} For a perfect conductor, {this reads}
\begin{equation}
\label{Eq. one-half}
  \mathcal E_1=\frac{1}{2} \,   \mathcal E_{1}(\mbox{parallel-plates})=-\frac{\hbar cA}{16\pi^{2}}.
\end{equation}
This result is independent of any internal length scale that characterizes the self-complementary geometry.  Thus, for example, the size of the strips,  the squares, or the wedges in Fig.~\ref{Fig-SelfDual} do{es} not enter the expression for the energy in the first reflection.

The higher-reflection corrections to Eq.~(\ref{Eq. one-half}) are no worse than the case of two parallel plates, where they are less than 8\%. This is because the higher reflection terms involve higher powers of the $\bbr$ of the self-complementary geometry.  The {absolute value of the} eigenvalues of $\bbr$ are all less than (or equal to) unity (whereas {those of }an infinite reflecting plate, are all unity),  so higher reflections' contributions are further suppressed.

If the separation between $\Sigma$ and the reflecting plate is much smaller than the length scale of the structure on the screen ({\it e.g.\/} $d\ll a$ in Fig.~\ref{Fig-SelfDual}), then our result follows from the {\it proximity force approximation} (PFA) and is not surprising \cite{Derjaguin56}. This approximation treats the objects locally as parallel plates and is exact in the limit of close proximity.  When the separation is comparable to or larger than the length scales of $\Sigma$, then our result is by no means obvious. In fact, one might have thought that a mesh like the checkerboard completely reflects electromagnetic waves with wavelength $\sim d$, when $d\gg a$ leading to $\mathcal E_{1} \to\mathcal E_{1}(\mbox{parallel-plates})$ as is the case, for example, for corrugated or pitted conductor when $d$ is large compared to the length scale of the corrugations or pitting \cite{Emig01}.

An interesting special case is {a}{ half-plate}. On dimensional grounds the energy can be separated into two contributions,  $\mathcal E=\mathcal E_{\rm area}+\mathcal E_{\rm edge}$, the first proportional to the area and the second proportional to the edge length.   The former is given by $\mathcal E_{\rm area}= -{\pi^2 \hbar c A}/{720 d^3}$ where $A$ is the area and $d$ is the separation distance.  Equation~(\ref{Eq. one-half}) rules out the edge term in the first reflection, $\mathcal E_{{\rm edge}\, 1}=0$, since the energy differs from parallel plates only by a trivial factor of one-half --- which accounts for the area of the {\it half}-plate. A similar argument shows that for scalar fields, the  Dirichlet and Neumann edge terms are equal in magnitude and opposite in sign in {the} first reflection approximation.  The absence of the edge term in the first reflection for a half-plate geometry was first observed in Ref.~\cite{Maghrebi10}, as a limiting case of a wedge with zero opening angle.

We believe this argument can be generalized further.  Consider a screen $\Sigma$ with structure characterized by a minimum length scale, $a$.  Then as the separation from a reflecting plate, $d$, goes to zero, the leading contribution to the Casimir interaction energy, proportional to $A/d^{3}$, is captured by the PFA, because the set up is locally indistinguishable from segments of parallel plates.  If the next term in the expansion in $1/d$ is also determined locally, then it must be   proportional to the perimeter of the screen $\Sigma$,
\begin{equation}
  \mathcal E \underset{d\to 0}{\sim}- {\hbar c} \left(\alpha_A \frac{A}{d^3}+\alpha_P \frac{P}{d^2}+\cdots\right).
\end{equation}
The coefficient of the first term is set by the PFA, $\alpha_A={\pi^2  }/{720 }$.   Locally as $d\to 0$ any edge looks like a half-plate so we can conclude  that $\alpha_P$  is zero in first reflection, which in turn implies that the edge effects are very small. More precisely, \emph{if edge effects are local}, for electromagnetism  $|\alpha_P|\ll \alpha_A$, while for Dirichlet and Neumann boundary conditions (for which, the edge term is not absent in the first reflection), we have $|\alpha_P|\sim \alpha_A$ \cite{Graham10, Maghrebi10}.  {The absence of an edge term in the electromagnetic Casimir energy for a piston geometry \cite{Hertzberg05} supports this conjecture.}

Finally we consider a screen with small apertures.  The classical application of the BP in optics is to relate diffraction by an aperture to diffraction by its complement.  If the aperture is small, then its complement is a small object whose Casimir interactions can be computed in the Casimir-Polder limit \cite{Casimir48-1}. The exact relation involves  interchanging polarizations as given in Eqs. (\ref{Eq. T-LL EM}) and (\ref{Eq. T-RL EM}).  For a screen with a small hole opposite an infinite reflecting plate {it} is  easy to show that in the first reflection approximation, the energy is given by the interaction of two infinite plates minus the interaction of the complementary object with the infinite plate.  This gives a small correction to the force between parallel plates.

An interesting case is the lateral force between two screens with perforations.  The energy can be written as the sum of the interaction between parallel plates (without the holes), the hole-plate and plate-hole interactions, and the hole-hole interaction, plus higher reflections,  $\mathcal E=\mathcal E_{{\rm plates}}+ \mathcal E_{\rm plate-hole}+\mathcal E_{\rm hole-plate}+\mathcal E_{\rm hole-hole}+\cdots$. Only the interaction between the holes can give rise to a lateral force between the two plates and $\mathcal E_{\rm hole-hole}$ is given by the  electromagnetic interaction between two objects of the same size and position as the holes.  If the holes are small compared to the interscreen separation the higher reflections are negligible.

\begin{figure}
 \centering
 \includegraphics[scale=.95]{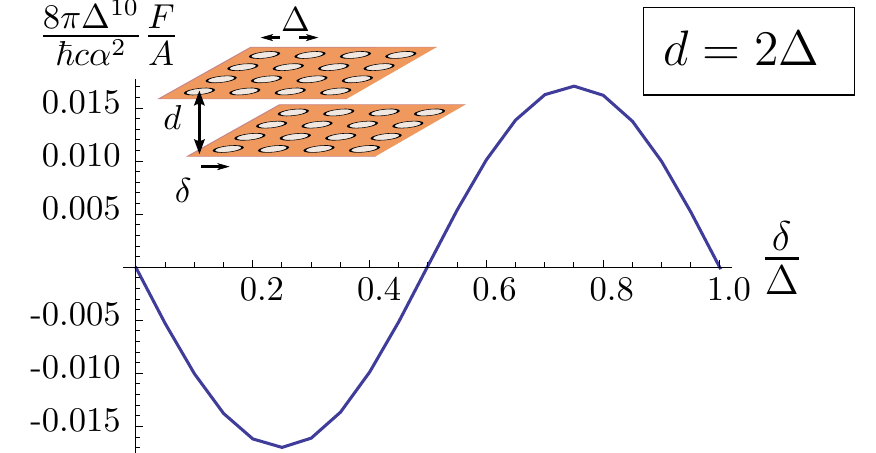}
 \caption{The lateral force {per area} as a function of the displacement $\delta$. For this graph, we chose $d=2\Delta$. The point $\delta=0$ is an stable equilibrium of this configuration.} \label{Fig-PerforatedPlates2}
\end{figure}

As an application of this, we consider two perforated plates, each with a square array of small circular holes of radius $R$ and center-to-center separation $\Delta$; the two plates are placed in parallel separated by a distance $d$.  According to the BP and in leading order in $R/d$ and $R/\Delta$, the lateral force between the two plates is identical to the lateral force between two arrays of discs with the same size and spacing as the holes.    The Casimir interaction between two objects with electric polarizability \footnote{For simplicity we ignore magnetic response.}  matrices $\alpha_1$ and $\alpha_2$ is given by  \cite{Feinberg70, Emig09}
\begin{equation}
  \mathcal E =-\frac{\hbar c }{8 \pi r^7}\left(13 \, {\rm tr}(\alpha_1\alpha_2)-56 \, {\rm tr}(\alpha_1 \alpha_2 \Omega)+63 \, {\rm tr}(\Omega \alpha_1 \Omega \alpha_2)\right).
\end{equation}
In this equation, $\Omega= \hat n \hat n^T$ with $\hat n$ being the unit vector connecting the two dipoles separated by a  distance $r$. The force between the two arrays of dipoles {can be computed easily}.  For a perfect conductor $\alpha_1=\alpha_2=\mbox{diag}\,\{\alpha, \alpha, 0\}$
  where $\alpha=4 R^3 /3\pi$; only the polarizability components parallel to the disk are nonzero. Figure \ref{Fig-PerforatedPlates2} shows the lateral force as a function of the lateral displacement $\delta$.

Finally we comment on the physical conditions that must be satisfied for the BP to apply to a screen made of a good conductor like gold.  On the one hand
the screen must be {a good conductor} . On the other hand, it should be thin enough to be considered as a screen of negligible thickness. So the thickness $t$ should satisfy $\delta \ll t \ll d$ where $\delta$ is the skin depth and $d$ is the separation distance. {The skin depth} is determined from the frequency by $\delta=\sqrt{2/\mu_0 \omega \sigma}$. For separation $d$, the relevant frequency is $\omega \sim 2\pi c/d$, and the inequalities become  $\sqrt{ {d/\pi c\mu_0\sigma}} \ll t \ll d$. For gold the skin depth is $\delta \approx 5$nm at $d\approx 1\mu$m and decreases proportional to $\sqrt{d}$ as $d$ decreases.  For measurements in the $d\sim 0.5 - 1.0\mu$m range a thickness of 100nm should suffice{ while for} $d\sim 0.1 - 0.5\mu$m, {it should }{be reduced to }${t \sim} 30$nm.

We thank T.~Emig, N.~Graham, M.~Kardar and M.~Kr{\"{u}}ger for helpful
conversations.  This work was supported in part by the U.\ S.\ Department of Energy under cooperative research agreement \#DF-FC02-94ER40818 (RLJ {and MFM}) and the NSF Grant No. DMR-08-03315, DARPA contract No. S-000354 (MFM).

\end{document}